# OUTBURST ACTIVITY IN COMETS: II. A MULTI-BAND PHOTOMETRIC MONITORING OF COMET 29P/SCHWASSMANN-WACHMANN 1

Josep M. Trigo-Rodríguez[1,2]
D. A. García-Hernández [3,4]
Albert Sánchez[5]
Juan Lacruz[6]
Björn J.R. Davidsson[7]
Diego Rodríguez[8]
Sensi Pastor[9]
and José A. de los Reyes[9].

1. Institute of Space Sciences (CSIC), Campus UAB, Facultat de Ciències, Torre C-5 pares, 2a pl., 08193 Bellaterra (Barcelona), Spain. E-mail: trigo@ieec.uab.es
2. Institut d'Estudis Espacials de Catalunya (IEEC), Ed. Nexus, Gran Capità 2-4, 08034 Barcelona, Spain
3. Instituto de Astrofísica de Canarias (IAC), E-38200 La Laguna, Tenerife, Spain.
4. Universidad de La Laguna (ULL), E-38205 La Laguna, Tenerife, Spain
5. Gualba Astronomical Observatory (MPC442), Barcelona, Spain.
6. La Cañada Observatory (MPC J87), Ávila, Spain.
7. Department of Physics and Astronomy, Uppsala University, Box 516, SE-75120, Uppsala, Sweden.
8. Guadarrama Observatory (MPC458), Madrid, Spain.
9. Observatorio Astronómico Municipal de Murcia (J76), La Murta, Murcia, Spain.

ABSTRACT: We have carried out a continuous multi-band photometric monitoring of the nuclear activity of comet 29P/Schwassmann-Wachmann 1 from 2008 to 2010. Our main aim has been to study the outburst mechanism on the basis of a follow-up of the photometric variations associated with the release of dust. We used a standardized method to obtain the 10 arc-sec nucleus photometry in the V, R, and I filters of the Johnson-Kron-Cousins system, being accurately calibrated with standard Landolt stars. Production of dust in the R and I bands during the 2010 Feb. 3 outburst has been also computed. We conclude that the massive ejection of large (optically-thin) particles from the surface at the time of the outburst is the triggering mechanism to produce the outburst. Ulterior sublimation of these ice-rich dust particles during the following days induces fragmentation, generating micrometer-sized grains that increase the dust spatial density to produce the outburst in the optical range due to scattering of sun light. The material leaving the nucleus adopts a fan-like dust feature, formed by micrometer-sized particles that are decaying in brightness as it evolved outwards. By analyzing the photometric signal measured in a standardized 10-arcsec aperture using the *Phase Dispersion Minimization* technique we have found a clear periodicity of 50 days. Remarkably, this value is also consistent with an outburst frequency of 7.4 outbursts/year deduced from the number of outbursts noticed during the effective observing time.

1. INTRODUCTION

Comet 29P/Schwassmann-Wachmann 1 (hereafter SW1) has been considered for long time the archetype of comets exhibiting unusual changes in their coma appearance and brightness (Sekanina, 1982). SW1 moves along a quasi-circular orbit with eccentricity $e$~0.044 and semimajor axis $a$~6 AU (Marsden & Williams, 2003). A key goal behind a photometric monitoring is to deduce the comet rotation period from the observed activity. The idea is not new, first attempts to do so with comet 29P were performed many years ago (see e.g. Whipple, 1980; Meech et al., 1993; Cabot et al., 1997). In any case, many difficulties arise in trying to understand 29P outbursts and its photometric behavior. During an outburst its nuclear magnitude typically increases by 2 or 5 magnitudes in a behavior that seems unusual, but that would be typical for primitive objects like e.g. Centaurs, Trans-Neptunian Objects (TNOs) located at the appropriated heliocentric distances. As we pointed out previously, being so far away from the Sun its surface temperature is below the sublimation of water ice so other physical processes are suspicious of playing a role for generating outbursts (Trigo-Rodríguez et al., 2009). In that previous paper that we will cite along the text as Paper I we presented a detailed 5-years photometric monitoring. Despite that such previous follow-up was mainly performed by relatively small telescopes, interesting conclusions on the activity of this Centaur were obtained, but other important issues on the mechanism producing these outbursts, and the periodicity of this activity remained open. Since the publication of Paper I we have continued our monitoring of this object with four main objectives: 1) To go a step forward in monitoring this object by using multi-band photometry in different Johnson-Kron-Cousin standard filters, 2) to use Landolt and Stetson calibration fields for accurately calibrate the photometric estimations in order to find out the accuracy of 10-arcsec nuclear photometry in the different filters, and 3) to identify the periodicity and plausible triggering mechanism producing these outbursts.

We were encouraged by Paper I results to continue our follow-up of SW1 activity to learn more on the processes that take place in the coma after an outburst. Due to the energy driven by explosive activity in 29P surface we previously suggested that mm-sized particles leave the surface and their progressive sublimation in the coma causes the peculiar coma evolution behavior described in Paper I. Our multiband monitoring confirms such an explanation, being consistent with the reflectivity variations observed in the coma of this comet after the outbursts. In Section 2 are compiled the observational data and reduction procedures taken into account in the present work. Section 3 we describe the outburst detected during the 2008-2010 observational period, some of them widely monitored in the different photometric bands. In Section 4 we discuss the implications of our observations for explaining cometary activity and dust processes taking place in the coma. Our final conclusions are summarized in Section 5.

2. OBSERVATIONAL DATA AND REDUCTION PROCEDURE.

We have performed multi-band CCD photometry from several amateur and professional observatories (Table 1). A log of the observatories used together with some relevant information such as telescope, instrument, resolution, and observers is shown in Table 1. Optical photometry in the Johnson V, R, and I filters was obtained with the IAC-80 telescope (Observatorio del Teide, Spain) equipped with the CAMELOT CCD (see e.g. http://www.iac.es/telescopes/iac80/CCD.htm for more details). The IAC-80 Johnson VRI magnitudes were derived by using standard aperture photometry tasks in *Image Reduction and Analysis Facility (IRAF)* software is distributed by the National Optical Astronomy Observatories, which is operated by the Association of Universities for Research in Astronomy, Inc., under cooperative agreement with the National Science Foundation.). The data reduction process basically consisted in the bias subtraction, flat-field correction, and flux calibration. This time the nuclear photometric accuracy given for a standard aperture of 10 arcsec has been improved in reference with Paper I by using Landolt, and Stetson calibration fields (Landolt, 1992; Stetson, 2000). The photometric reduction procedure for the small-sized telescopes was performed by using an additional software package called LAIA (Laboratory for Astronomical Image Analysis) successfully tested for obtaining high-precision stellar photometry (García-Melendo & Clement, 1997; Escolà-Sirisi et al., 2005). The procedure was similar to the described by Snodgrass et al. (2006), but limited to our 10 arcsec selected aperture. By using these standards we were able to quantify the typical data accuracy being better than 0.05 magnitudes in the different VRI bands. These results are clearly improving the 0.1-0.2 error in R magnitude obtained in Paper I. The spatial scale obtained by our instruments was between 0.3 to 2.3 arcsec/pixel which correspond respectively to a resolution of about 1,255 and 9,625 km/pixel at the Earth-comet distance $\Delta=5.77$ AU (see Table 1 for more details). These ideal values are typically affected by seeing so the effective resolution is between 1 to 2 arcsec or roughly about 6300 km/pixel. Our experience says that such resolution is good to follow the overall nuclear activity of this Centaur in the above mentioned standard aperture (Trigo-Rodríguez et al., 2009).

Table 1.

A total of 498 photometric measurements of SW1 in the different filters have been considered in the present paper, although our monitoring is still continuing. The data were obtained by 8 experienced observers on 217 nights extended between Jan. 2008 and June 2010. This monitoring can be considered complementary to the 5-years follow-up presented by our team in Paper I. The observational data obtained by our team are available to other cometary scientists under request.

3. OBSERVATIONAL RESULTS

The complete set of photometric observations obtained in the present work is shown in Figure 1. The data profile evidences a good coverage except for the periods when the comet was in conjunction with the Sun. We describe below some of the most important features that were detected in our monitoring of the outbursts.

Following the criterion described in Paper I where we counted as outburst any increase in one magnitude in the coma brightness we can identify the main bursts in the photometric profile. Taking this criterion into account our data reveals 11 outbursts from Jan. 2008 to Feb. 2010 (see Fig. 1). Due to weather circumstances some of them are better covered than others. The different outbursts detected, and several parameters to describe their particular strength are included in Table 2. Note that the table not only shows the increase in magnitude, but also the amplitude interval among consecutive outbursts. We found that a significant number of outbursts are occurring after 20-40 days.

In Fig. 1 we evidence two main types of outbursts that we would basically call: single and multiple. Single outbursts are exhibiting a progressive increase and decrease in the coma magnitude. Good examples of single outburst are those labeled on Table 1 as A, C, H, J, K, and L. In these cases, the activity increases progressively in about 20 or 30 days. The multiple outburst behavior is exemplified by the sequence of outbursts listed from D to G. A very simple explanation is behind of this behavior: the presence of multiple active regions that, as the nucleus rotates, are exposed periodically to solar radiation. More details are presented in the discussion, but note that a nucleus with several active regions rotating every 50 days can explain the photometric behavior.

The period between JD 2454800-2454960 is shown in a separate window in Fig. 2a. We remark that the activity profile starts to grow progressively from a quiescent period in magnitude +15.8 R until reaching a peak called D in +12.5 R on JD 2454822.7. All the upwards profile is characterized by high I magnitude values. Only 12 days after that, just on JD 2454834.6, is noticed peak E with similar strength. Note that the two peaks' pattern reappears on JD 2454871.4, and JD 2454883.4 (peaks F and G) with a similar spacing of 12 days, but in this case much better sampled. Note that both peaks are repeated after 50 days. Curiously, despite the good data sampling, this pattern disappears because the following outburst occurs on JD 2454944.3.

Our best monitored outburst took place around Feb. 2, 2010 and has been included in full detail in Fig. 2b. We posted an alert to the astronomical community on this unusually bright outburst in CBET #2160 (Trigo-Rodríguez et al., 2010). On Feb. 3.18 first multi-band photometry was obtained from Gualba Observatory (MPC442) that shows the comet exhibiting the bright core characteristic of a massive outburst (Fig. 3). At that time the photometry of the 10-arcsec standard aperture shows an infrared increase that makes the nuclear magnitude brighter in I than in R, or V bands. The bright dust globule formed by the release of dust particles is evolving outwards, as confirmed by high-resolution additional observations taken from IAC80 (Fig. 4).

4. DISCUSSION.

4.1. THE PERIODICITY OF 29P OUTBURSTS.

To better understand the physical processes behind the production of outbursts by comet 29P, its rotation period is a key parameter that needs to be constrained. Different researchers have been trying this, but scientific literature shows large discrepancies. We think that the most probable origin of these discrepancies are insufficient or biased data sampling. Just to cite some examples of published period data in order of decreasing value, Stansberry et al.(2004) reported a rotation period of about 60 days or more, while Trigo-Rodríguez et al. (2009) and Moreno (2009) found about 50 days emphasizing the need of more precise observational data, and finally shorter periods were found by Jewitt (1990) 6 days, and Whipple (1980) 4.97 days. Meech et al. (1993) suggested that the nucleus exhibits a complex spin state with two really short periods of 14 and 32 hours. Obviously, the characterization of the spin rate of 29P nucleus from observational data requires a good coverage, but also having very accurate observational data. In this regard, the more accurate photometric accuracy (at least 0.05 mag.) of the 2008-2010 data here presented in comparison with our previous coverage presented in Paper I allows to better study the existence of periodicity in the photometric profile. Because the data sampling is unevenly spaced, we used different tools in the periodogram analysis, including precise determination of minimums by the same software AVE that we used previously (Trigo-Rodríguez et al., 2009). Basically we used the Phase Dispersion Minimization (PDM) (Stellingwerf, 1978) and the Discrete Fourier Transform (Deeming, 1975) methods. As the results of both methods are similar, just for simplicity we discuss here only PDM periodograms.

The PDM periodograms appear in Fig. 5. For discussion purposes of the values given in the literature we have just separated the periodogram in two windows, one for short periods of less than 10 days (Fig. 5a), and other for periods until 100 days. We note that there is not evidence of periodicity in the signal in Fig. 5a, except for the period of 1 day that is a an alias produced by the typical 24-h observational window. This periodogram is so clean that demonstrates that the comet's activity during 2008-2010 was not hiding a periodicity lower than ten days. In other words, if the profile activity is a reflex of active regions in the nucleus, its rotation was not in the plotted periodicity window. Fortunately, the second periodogram shown in Fig. 5b includes additional clues with three clear peaks over the noise. The most obvious is 66.94±0.03 days, followed by other at 55.21±0.07 days, and a third one at 50.02±0.07 days. The first value of about 67 days is just an alias (i.e. period combination) for the synodic period of about 392 days as we also found in Paper I. This idea is reinforced by the fact that additional exploration of larger periodicities also shown synodic period aliases for 128 and 200 days. Consequently, we think that the 50 days period can be given as a temptative value for the rotation period of this comet. In any case, we should be cautious when identifying the outburst period with the rotational period. The two may be related, but the former could be a low-number multiple of the latter because the number of rotational periods between outbursts could also vary stochastically, as apparently is also the case for 9P/Tempel 1 (Belton et al., 2008). Moreover, such a large period seems to be unusual among the population of TNOs and Centaurs that are exhibiting rotation periods typically lower than one day (Sheppard et al., 2008), and also for other comets (Samarasinha et al., 2004). Note also that we have not constructed and compared our observational results with the theoretical rotationally phased light curve as it is quite difficult for this object. It is obvious that the intensity of the outbursts is highly variable and also depends of the reliability of the observational coverage. Despite of this, taking into account the two periods where the comet was in solar conjunction (about 263 days without observations in our sample), the average outburst frequency during the effective observing period was 49 days. Note that this value is again consistent with the deduced rotation period from periodograms. Such frequency corresponds to 7.4 outbursts/year that is almost identical to the value previously found by our team (Trigo-Rodríguez et al., 2009). Note that this periodicity might be directly related with the rotational period of this comet, but we cannot rule out that it was just a multiple of such a period. The spurious (less defined) secondary peak at about 55 days would be consequence of additional active areas exhibiting sporadic activity, or other irregularities caused by stochastic evolution of the surface or subsurface properties.

### 4.2. QUANTIFYING THE DUST PRODUCTION: AF$\rho$

In previous sections we have shown that our observations suggest several active areas producing dust in the surface of this Centaur. The scenario seems to be complex, and highly variable as a function of time. For example, the observed activity pattern during Dec. 2008 and May 2009 plotted in Fig. 2a can be explained by the presence of two regions becoming active every 50 days. Although the data sampling is not complete after peak G, it seems that the double peak pattern disappears after it. Where the active regions collapsing (under rubble) or was the activity damped for other reason? It seems from the important variations in the active regions with time that the different strength of each outburst could be consequence of resurfacing, e.g. some released materials are participating in covering the active regions. In a similar way, the creation of new active areas can be also explained in the context of tensile strength variations in the surface as consequence of the crystallization of amorphous ice (Prialnik, 2002; Trigo-Rodríguez & Blum, 2009). Consequently, a long time processing like this pictured before might explain some of the features observed in several comets, particularly the layering observed in comet 9P/Tempel 1 studied by Deep Impact spacecraft (Belton et al., 2007). However, without clear data on the rotational axis, we think that is not possible to discard another possibilities associated with a changing tilt geometry. For example, if the rotation axis was complex, some regions would be exposed to sun light from time to time. Perhaps this is not so likely taking into account the 50 days periodicity behind most of the outbursts.

We suggested in Paper I that the source of the energy driving cometary outbursts would be associated with the explosive sublimation of ice-rich regions that, after a massive break up of the surface, are exposed to solar radiation (Trigo-Rodríguez et al., 2009). This massive release of material are produced from active regions of the surface. As bigger is the region, more massive would be the release of material participating in the outburst. The variability would be then explained by the formation of a rubble mantle (see e.g. Jewitt, 2008), perhaps due to collapse of the region when volatiles-depletion reduces the tensile strength of the surface. Other possibility is the accumulation of those large particles that cannot escape the nucleus. Stanberry et al. (2004) used thermal models fitted to photometry at 8, 24 and 70 μm in Spitzer observations to estimate a nuclear radius of 27±5 km. The escape velocity ($V_{esc}$) is defined as the minimum velocity that a grain needs to win the gravitatory attraction of the nucleus. It is calculated from the well-known equation:

$$V_{esc} = \sqrt{\frac{2\,G \cdot M_c}{R_c}} \qquad [1]$$

By considering a comet radius of 27 km, and assuming a spherical shape for the nucleus with a bulk density of 1000 kg/m$^3$, we have derived $V_{esc} \approx 20$ m/s. In Paper I we found that a 1mm diameter meteoroid with an assumed bulk density of 500 kg/m$^3$ is leaving typically the nucleus at $V_{ej} \approx 22$ m/s. We note that $V_{ej} \propto m^{-1/6}$ and, consequently, the impulse for large meteoroids decreases significantly. For example, a 1-cm diameter particle with the same density would have a $V_{ej} \approx 7$ m/s that is clearly inferior to the escape velocity for this comet. Consequently, the nucleus gravity creates a cutoff for large particles. However, we should remark that the 5-100-μm particles escaping SW1 nucleus are virtually unobservable in cometary comae at visual wavelengths. However, we want to remark that the previous simplistic approach is only qualitative because the escape velocity is really only relevant if gravity is the only force acting on the dust grains. In the current problem, we expect two other strong forces at play - gas drag and radiation pressure - that may change the picture dramatically.

Obviously μm-sized particles in the coma are evolving much faster under the effect of gas drag and radiation pressure. By using the resolved coma images obtained with the Pic du Midi 1-m reflector kindly provided by F. Colas on Feb. 4 and 5, and those obtained from IAC80 and other instruments (Table 3) we have found an ejection velocity of 250± 80 m/s that is smaller than the value obtained by Colas and Maquet (2010). This result suggests that the outburst probably occurred around Feb. 2.30.

Information about dust production for the different particles sizes can be obtained. In order to discuss the rate of activity of comets is usually measured the rate of activity of the nucleus $Af\rho$, typically given in cm. This parameter defined by A'Hearn et al. (1984) is a measure of the dust production rate, which does not attempt to disentangle the effects of albedo $A$, filling factor $f$ and aperture $\rho$, hence allowing direct comparison between data sets obtained with different instruments. This parameter can be computed from the following equation given by A'Hearn et al. (1984):

$$Af\rho = \left(\frac{(2\Delta r)^2}{\rho}\right)\frac{F_{com}}{F_\odot} \qquad [2]$$

where $\Delta$ and $r$ are the geocentric and heliocentric distances in AU; $\rho$ is the diameter of the photometric aperture in km, and $F_{com}$ and $F_o$ are the cometary and solar fluxes in the respective filters. The measured $Af\rho$ values for R and I filters are shown in Fig. 6. In a similar way that we noted how the I magnitude increases during an outburst, it is clearly visible in Fig. 6 that the production of large grains after the 2010 Feb. 3 outburst is announced by a significant increase in the in $Af\rho$ value in the I band. Going back to quiet activity the $Af\rho$ values in the R and I bands tend to become similar, like in the data before the outburst. Consequently, it seems that such behavior in the I band could be used to predict outbursts in the future.

Recently the unexpected second massive outburst of comet 17P/Holmes provided interesting clues on the nature of these outbursts. For example, it was previously proposed that the scale of 17P outburst could be explained by the disintegration of one of such outer layers, which after the separation from the surface would have broken apart into the micrometer-sized dust observed in the coma (Sekanina, 2007; Thomas et al., 2007). Dello Russo et al. (2008) found that the spatial distributions of measured volatiles in the coma are consistent among them and that only a minor contribution from sublimating icy grains was identified in their observing aperture. We interpret those observations as for most volatiles being produced by the explosive sublimation of the ice-rich surface at a very constrained timescale. The largest particles became fine dust during the first stages of expansion, and only few particles contributed to enhance the amount of volatile in the coma due to additional sublimation in the latter days.

### 4.3. THE EVOLUTION OF 29P OUTBURSTS

Massive release of particles triggering the outbursts seems to be typically produced from active areas in the surface. Two plausible scenarios are 1) that the volatile-rich areas are activated when the Sun rises in the local horizon or 2) the outbursts are produced by a sub-surface process that takes place at any local hour (due to heat conduction effects). Whatever the process is a massive release of particles is produced during the outburst. Large, and fresh particles reaching the coma are efficiently heated by the sun so their sublimation starts. Then, the progressive fragmentation of these particles takes place in a second stage of the outburst, increasing the number density of micrometer-sized dust, and consequently increasing the brightness in R, and V. The particles are released forming a fan, but later the coma become pseudo-spherical (globule-like) as the fine dust expands outwards (see e.g. Fig. 3 and 4). By doing so, the surface brightness of the coma decreases as we expect as consequence of decreasing the number spatial density of dust in the line of sight. An outburst is announced by a sudden increase in the comet magnitude produced by massive injection of particles to the coma from an active region. Ulterior sublimation of the ice-rich dust produces a progressive increase in the spatial density of micrometer-sized dust that starts to reflect light to produce the outburst in the optical range.

On the other hand, our results indicate that 29P outbursts are periodic so are necessarily produced by nuclear activity. Consequently, it is not needed to invoke the impact of meteoroid-sized bodies against cometary nucleus to explain them (Gronkowski, P., 2004). Of course, in statistically limited cases, such type of event would be behind the excavation of the surfaces of dormant comets, acting as a triggering mechanism. In any case, the fascinating behavior of comet 29P suggest that it is a pristine object that would be a nice target of exploration by future space missions.

5. CONCLUSIONS.

We have presented an unprecedented effort to monitor the photometric behavior of comet 29P using different Johnson-Kron-Cousin filters. As a by product of this monitoring we have been able to infer information on the development, and nature of these processes. To summarize our findings:

- a) We have demonstrated that a multi-band monitoring of 29P activity can be used to study the physic processes behind the development of cometary outbursts. Future high-resolution spectroscopic observations before, during and after the outbursts are encouraged. Such observations would help, particularly, to better understand the processes involved in the sublimation and fragmentation of the particles that we propose here.
- b) Finally, we think that we have been able to constrain the rotation period of 29P nucleus from photometric data. PDM periodograms show a 50 days periodicity that we consider a temptative value for the rotation period of this Centaur that fits most of the outbursts recorded from 2002-2010. Such periodicity is also corroborated by the photometric profile produced by the presence of two active regions in 29P surface in the period Dec. 2008-Feb. 2009.
- c) The average outburst frequency during the full monitored period was 49 days. Note that this value is consistent with the deduced rotation period in *b)*. Such frequency corresponds to 7.4 outbursts/year that is very close to the value previously found by our team (Trigo-Rodríguez et al., 2008). In consequence, outburst activity of Centaur 29P is periodic, so the triggering mechanism for outbursts is the periodic insolation of an active region.
- d) On 2010 Feb. 2.3 took place the brightest outburst observed during the last years. Computed Afρ values (41,000 cm) in the R band well exemplifies the energy driven by these outbursts. A follow up of the particles by using resolved images of the expanding coma indicate that the particles were ejected with a velocity of 250± 80 m/s.

ACKNOWLEDGEMENTS

We thank the valuable comments, and encouragement received from Dr. A. Fitzsimmons (Queen's University Belfast). We also appreciate a detailed and very constructive review received from Dr. Jacques Crovisier. The authors thank *Grup d'Estudis Astronòmics (GEA)* for providing the reduction software used in the present studies, very especially LAIA and AVE programs developed by R. Barberà and J.A. Cano. Part of this work is based on observations made with the IAC-80 telescope under the Spanish Instituto de Astrofísica de Canarias CAT Service Time. The IAC-80 is operated by the Instituto de Astrofísica de Canarias in the Observatorio del Teide. D.A.G.H. acknowledges support for this work provided by the Spanish MICINN under the 2008 Juan de La Cierva Program and under grant AYA-2007-64748.

# TABLES

| Observatory | MPC code | Telescope | Resolution (arcsec) | Long (º) | Lat (º) | Observer |
|---|---|---|---|---|---|---|
| Gualba, Barcelona | 442 | SC 36.0 f/7 | 1.6 | 2º 31' 15" E | 41º 43' 16" N | A. Sánchez |
| Guadarrama, Madrid | 458 | SC 25 f/6.3 | 2.3 | 4º 01' 20" W | 40º 38' 14" N | D. Rodríguez |
| IAC80, Teide, Tenerife | 954 | C 82 f/11.3 | 0.5 | 16º 30' 33.4" E | 28º 17' 53.6" N | D. A. García-Hernández et al. |
| La Cañada Observatory | J87 | RC 40 f/10 | 1.2 | 4º 29' 30" W | 40º 36' 18" N | J. Lacruz |
| Montseny Astronomical Observatory, Girona | B06 | N 18 f/6 | 1.5 | 2º 31' 14" E | 41º 43' 17" N | J.M. Trigo-Rodríguez |

Table 1. Observatories, telescopes, resolution and people involved.

| Outburst date | Julian date | Peak label | Observed peak R magnitude (mag) | Magnitude increase (mag) | Peak semi amplitude (days) | |
|---|---|---|---|---|---|---|
| | | | | | UP | DOWN |
| Jan. 14.8, 2008 | 2454480.4 | A | +12.4 | 3.8 | - | 39 |
| Mar. 13.8, 2008 | 2454539.3 | B | +14.2 | 2.0 | - | 23 |
| Sep. 25.1, 2008 | 2454734.6 | C | +11.8 | 4.1 | - | 57 |
| Dec. 23.2, 2008 | 2454822.7 | D | +12.5 | 3.3 | 16 | - |
| Jan. 3.1, 2009 | 2454834.5 | E | +12.6 | 1.0 | - | 24 |
| Feb. 8.8, 2009 | 2454871.4 | F | +12.7 | 2.7 | 13 | 10 |
| Feb. 21.7, 2009 | 2454884.3 | G | +13.3 | 1.3 | 4 | 31 |
| Apr. 22.8, 2009 | 2454944.3 | H | +14.8 | 1.4 | 30 | (23) |
| Sep. 23.1, 2009 | (2455097.6) | I | (+13.9) | (2.0) | - | 20 |
| Nov. 10.2, 2009 | 2455145.6 | J | +13.5 | 2.4 | 28 | 40 |
| Feb. 3.2, 2010 | 2455230.7 | K | +11.7 | 4.3 | 46 | - |
| Apr. 16.8, 2010 | 2455303.3 | L | +12.7 | 3.5 | 30 | (19) |

Table 2. Specific details of the detected outbursts. The peak amplitude denotes the approximate time required for the comet to reach the maximum brightness (up) or be back (down) into a quiescent stage. This value is only given for those cases in which the outburst is well covered. Temptative values in case of poor coverage are given between parenthesis.

| Date (JD) | Observatory | Apparent Coma diameter (arcsec) | Coma diameter ($\times 10^3$ km) |
|---|---|---|---|
| 2455230.68 | 442 | 19.2 | 73±7 |
| 2455231.43 | 458 | 23.5 | 86±9 |
| 2455232.43 | 458 | 30.6 | 93±9 |
| 2455232.48 | IAC80 | 27.4 | 103±10 |
| 2455233.49 | 442 | 30.4 | 114±7 |
| 2455233.58 | IAC80 | 32.2 | 121±12 |
| 2455234.47 | 458 | 42.3 | 160±15 |
| 2455238.69 | 442 | 49.6 | 186±16 |

Table 3. Selected measurements of the coma diameter of SW1 during the massive outburst experienced in Feb. 2, 2010. For more details see the text.

# FIGURES

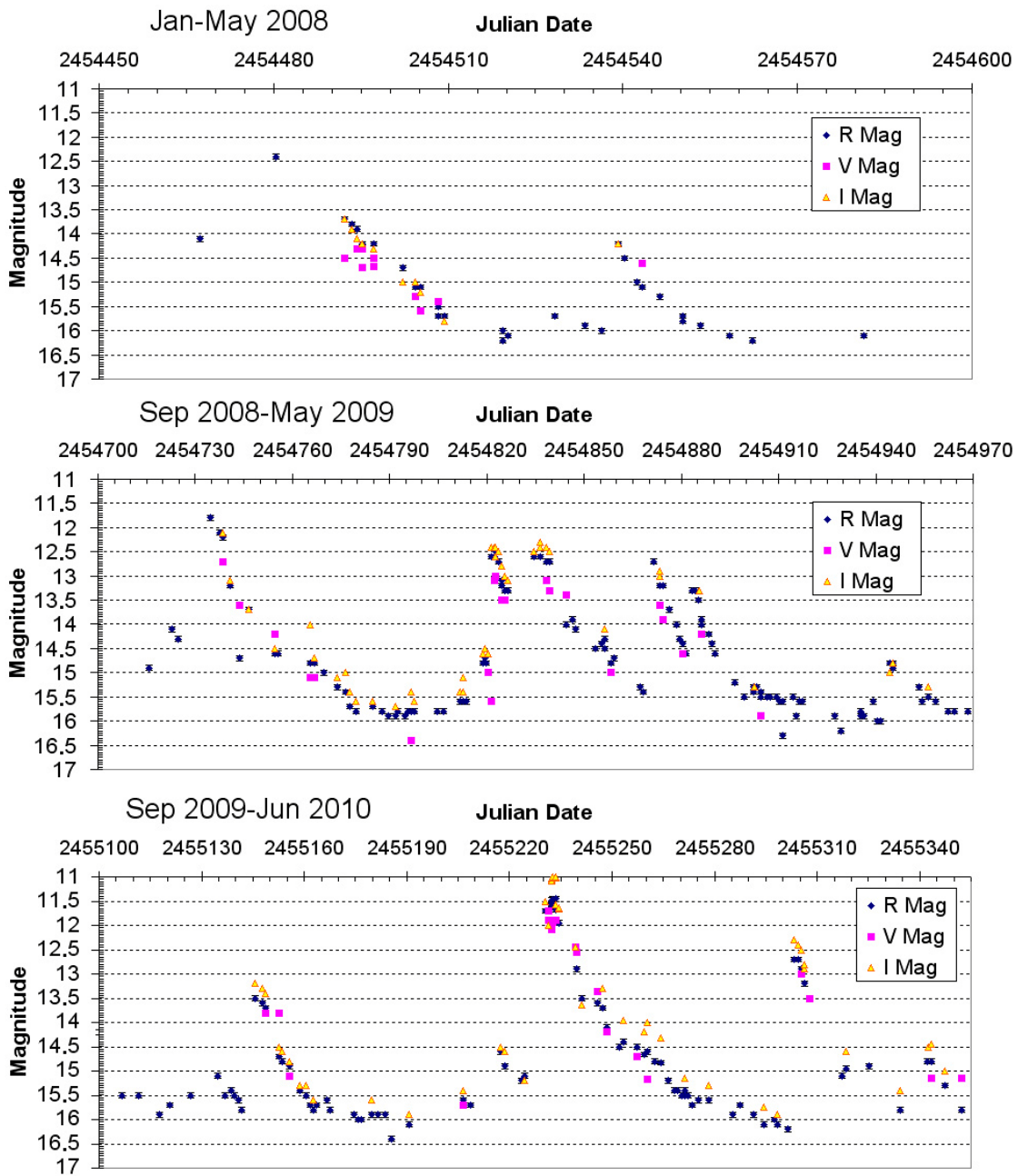

Figure 1. Photometric coverage using the different Johnson-Cousin filters during the monitored period. Error bars are not shown at the present image resolution, but magnitude accuracy was found to be better than 0.05 magnitudes.

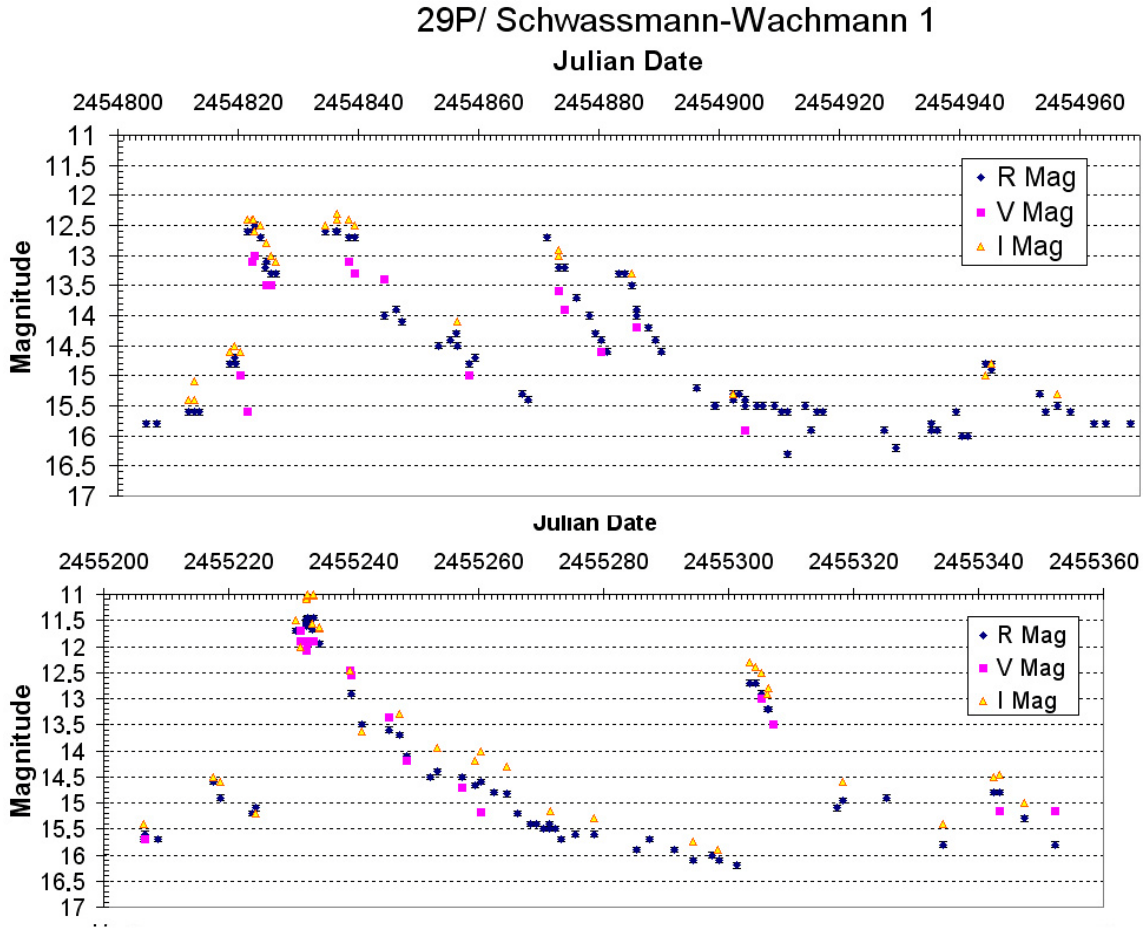

Figure 2. Detailed windows of the photometric profile showing the best sampled outbursts plotted in Fig. 1. a) Photometric curve in the different filters showing the double-peak behavior exhibited from Dec. 2008 to May 2009. b) Detailed photometry around the Feb. 2, 2010 outburst where are shown the observations made during the first **semester** of 2010. Note that the error bars (0.05 magnitudes) are not visible at this scale.

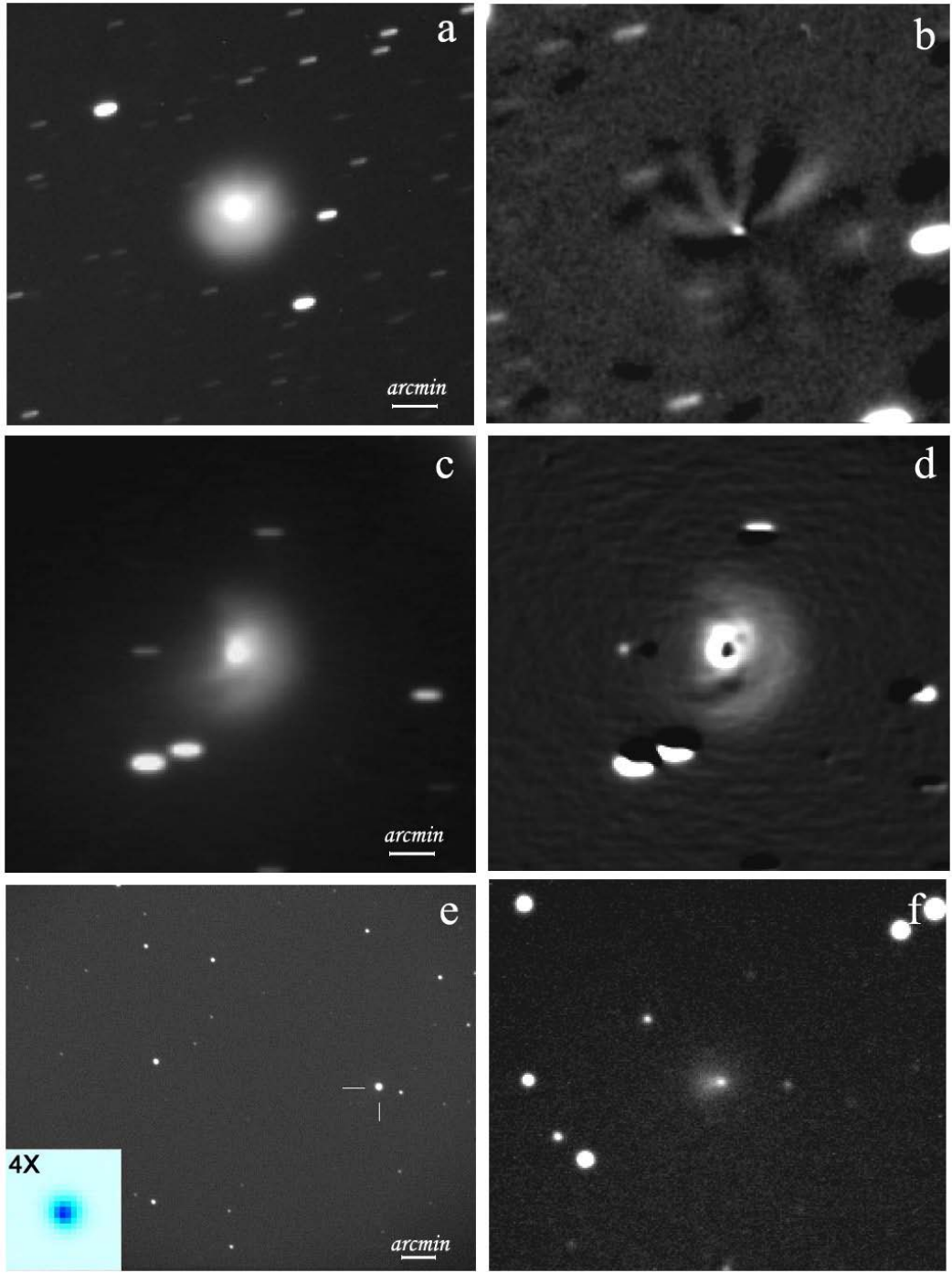

Figure 3. Some images exemplifying the changing appearance of comet 29P. a) A 42-minutes stack image obtained about two weeks after outburst C (see Table 2 for further details). Taken from MPCJ87 observatory by J. Lacruz on JD2454743.6. b) Larson-Sekanina filtered image of *a* (rotated by 15 degrees) evidencing the multiple jets that formed the fan at the previous image. c) 45-minutes stack image taken from J87 on JD 2454877.4, just one week after outburst F. d) Larson-Sekanina filtered image of *c* rotated by 15 degrees and shifted by 0.5 pixels to better see the fan structure. e) stellar appearance of 29P just after the outburst K when the comet was in +11.7 R magnitude. 3-minutes stack image taken from MPC442 by A. Sánchez on JD 2455231.5. f) Pre-outburst image of *e*, taken with the same instrument from MPC442 on JD2455223.7 when the comet was in +15.2 R magnitude. Note that the scale in the right-side pictures is identical to those shown at left.

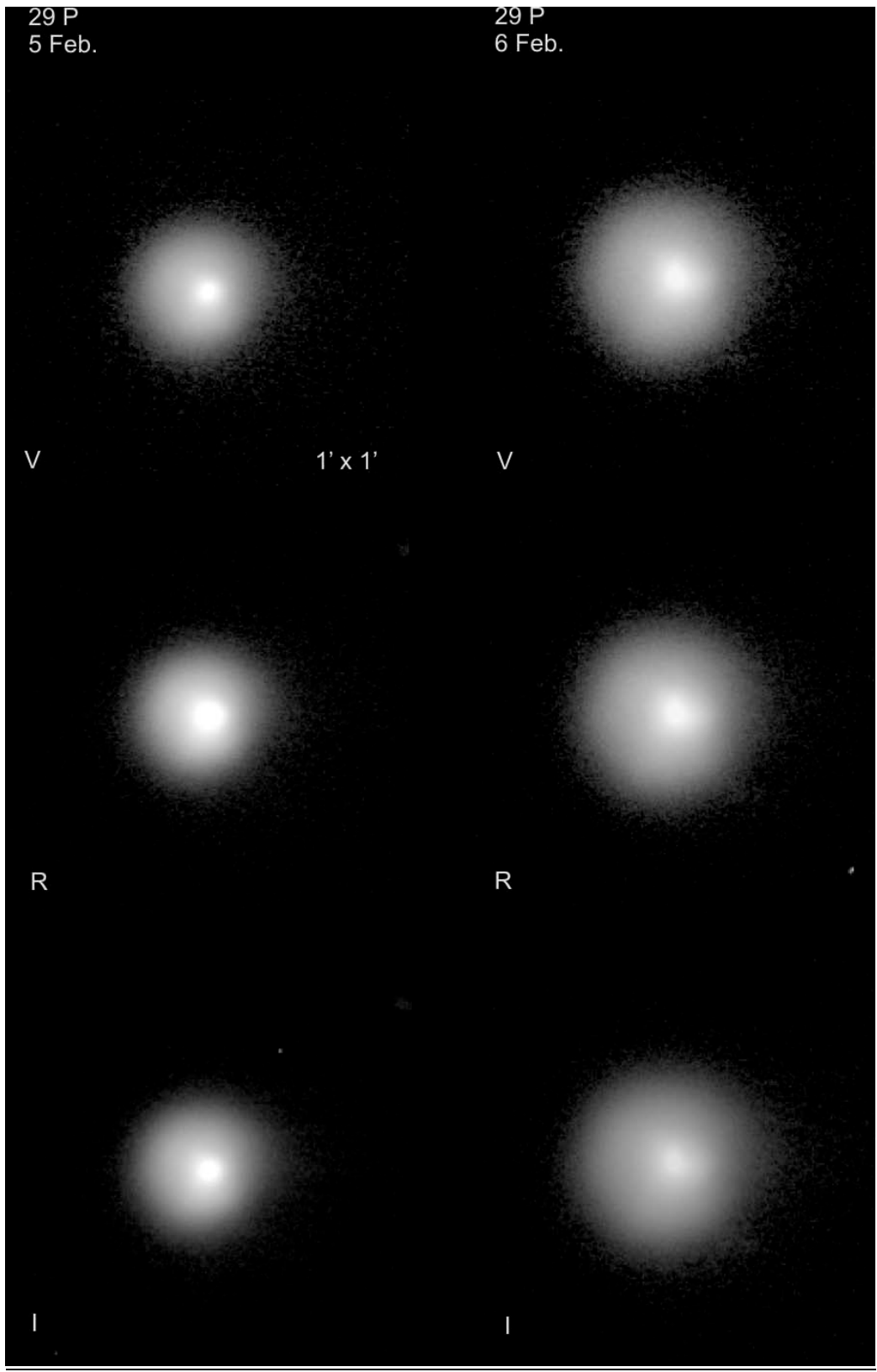

Figure 4. The expanding coma after the outburst as imaged with the IAC80 telescope with the different VRI filters for Feb 5.09 (left images) and Feb. 6.08 (right ones). The images are shown by using a logarithmic scale.

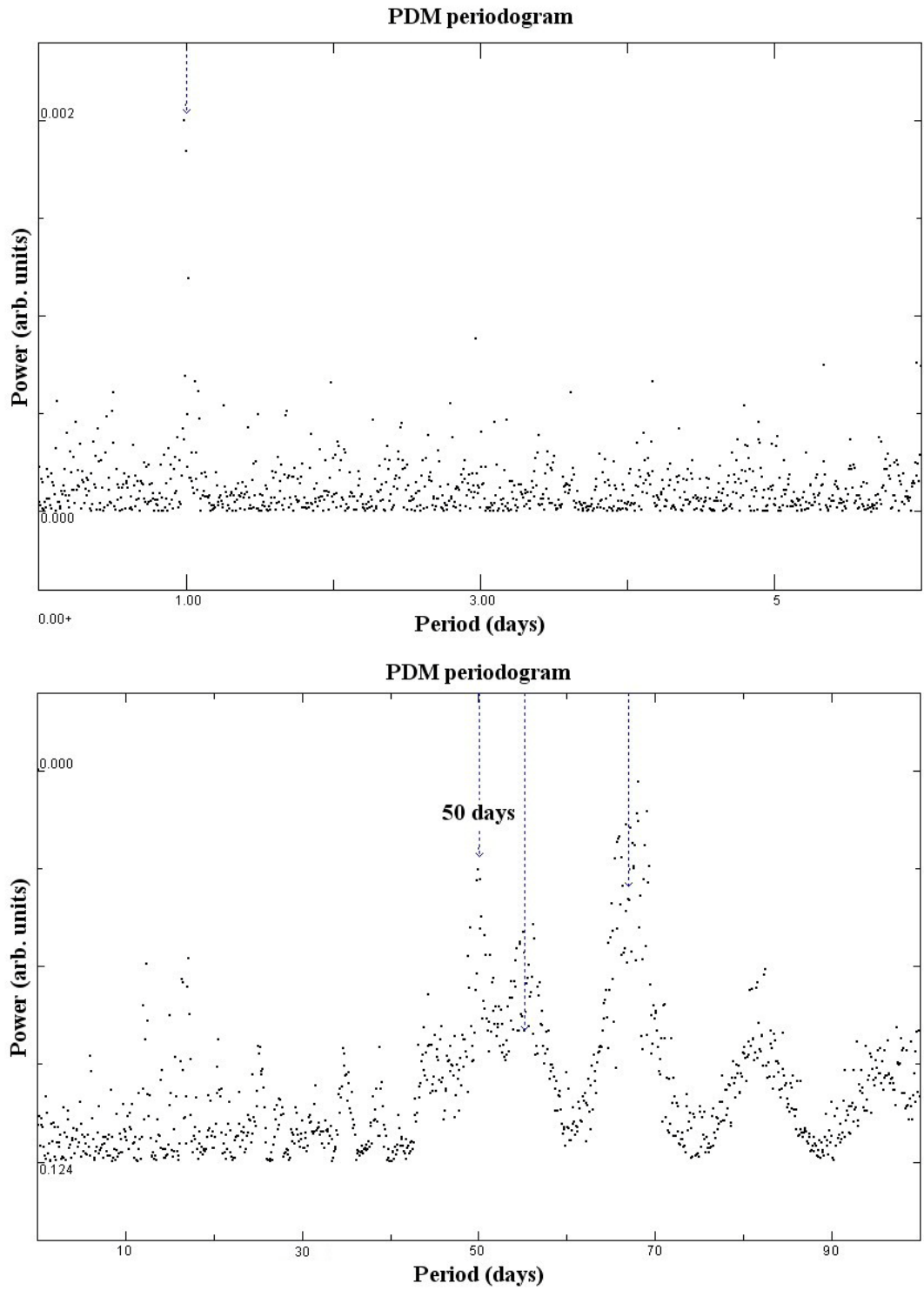

Figure 5. The periods shown by Phase Dispersion Minimization periodograms. The peak for a nucleus periodicity P=50.02±0.07 days is shown. For more details on the other peaks see the text.

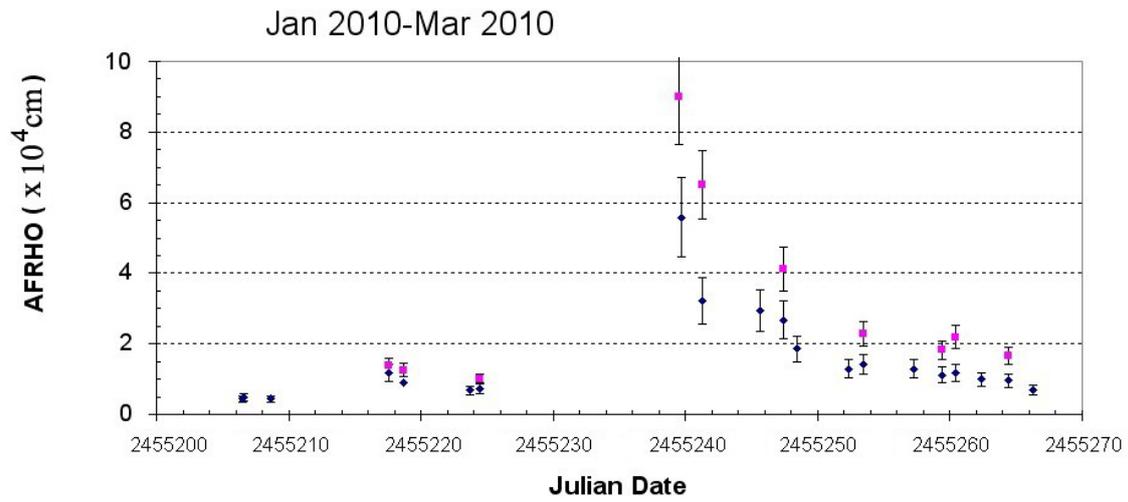

Figure 6. Computed *Afρ* values for the main outburst detected inside the January to March 2010 period. Blue dots are computed from R photometric band data, and pink squares for the I band.